# Surrogate analysis of volatility series from long-range correlated noise

Radhakrishnan Nagarajan

**Abstract:**

Detrended fluctuation analysis (DFA) [1] of volatility series has been proposed to identify possible nonlinear/multifractal signatures in the given empirical sample [2-4]. Long-range volatility correlation can be an outcome of static as well as dynamical nonlinearity. In order to argue in favor of dynamical nonlinearity, surrogate testing is used in conjunction with volatility analysis [2-4]. In this brief communication, surrogate testing of volatility series from long-range correlated monofractal noise and their static, invertible nonlinear transforms is investigated. Long-range correlated noise is generated from FARIMA (0, d, 0), with Gaussian and non-Gaussian innovations. We show significant deviation in the scaling behavior between the empirical sample and the surrogate counterpart at large time-scales in the case of FARIMA (0, d, 0) with non-Gaussian innovations whereas no such discrepancy was observed in the case of Gaussian innovations. The results encourage cautious interpretation of surrogate analysis of volatility series in the presence of non-Gaussian innovations.

**Author for Correspondence**

Radhakrishnan Nagarajan
Institute on Aging, UAMS
629 Jack Stephens Drive, Room: 3105
Little Rock, AR 72205, USA
Email: nagarajanradhakrish@uams.edu

# 1. Introduction

Detrended fluctuation analysis (DFA) [1] has been widely used to determine the scaling behavior of synthetic and natural processes. Subsequently, several extensions of the DFA have been proposed to understand the nature of correlations from the given empirical sample [2-6]. More importantly, long-range correlation in the magnitude series or volatility series has been proposed to identify possible nonlinear/multifractal signatures [2-4] in the given empirical sample. Long-range volatility correlations can be outcome of static as well as dynamical nonlinearity. While dynamical nonlinearities represent feedback processed inherent to the dynamics, static nonlinearities are an outcome of transfer function of a measurement device that maps underlying dynamics onto the given empirical sample. In order to make a finer classification on the nature of nonlinearity giving rise to long-range volatility correlation, surrogate testing is often used in conjunction with traditional volatility analysis.

In the present study, volatility series generated from FARIMA (fractional integrated moving average) models with Gaussian, non-Gaussian innovations and their static invertible nonlinear transforms is investigated. It is shown that volatility analysis in conjunction with surrogate algorithms, namely: Phase-randomized (FT) [7], amplitude adjusted Fourier transform (AAFT) [7] and iterated amplitude adjusted Fourier transform (IAAFT) [8] can be useful in identifying the nature of the process for empirical samples generated from FARIMA (0, d, 0) with Gaussian innovations and their static, invertible nonlinear transforms. However, their limitations are clearly drawn in the presence of non-Gaussian innovations. This is reflected by significant discrepancy in the scaling of the volatility series between the empirical samples and their surrogate realizations at considerably large time scales in the case of non-Gaussian. The present study is in conjunction with our recent investigation on the impact of non-Gaussian innovations on surrogate testing [9].

## 2. Methods

### 2.1 FARIMA (Fractional Integrated Moving Average) models

Classical modeling of stationary linear process relies on Wold decomposition theorem [10]. This in turn has given rise to the family of models termed as ARMA (auto-regressive moving average) consisting of a deterministic auto-regressive part (AR) and a non-deterministic moving-average part (MA). Stationary ARMA process $(X_t)$ is represented by the expression

$$\boldsymbol{f}(B)X_t = \boldsymbol{y}(B)\in_t \tag{1}$$

where $B$ represents the backshift operator such that $BX_t = X_{t-1}$; $\boldsymbol{f}(B)$ and $\boldsymbol{y}(B)$ are polynomials of order $p$ and $q$ with roots outside the unit circle; $\in_t$ is a zero-mean finite variance i.i.d process. In the present study, we consider $\in_t$ sampled from Gaussian and non-Gaussian i.i.d processes. ARIMA (p, d, q) (auto-regressive integrated moving average) is a generalization of (1) where $X_t$ is replaced by its $d^{th}$ difference, i.e. $\nabla^d X_t = (1-B)^d X_t$ for integer values $d$.

It is important to note that the above models are markovian in nature and can be useful in modeling short-term correlations. However, there have been several instances where the given empirical sample exhibits non-markovian or long-range correlated behavior. Empirical samples that exhibit long-range correlations can be modeled as fractional ARIMA or FARIMA (p, d, q) models. FARIMA (0, d, 0) is a special case of FARIMA (p, d, q) whose behavior is solely governed by the parameter $d$. The expression for FARIMA (0, d, 0) [11-13], (proposition 2.2 [12]) is

$$X_t = \sum_{i=0}^{M=\infty} r_i \in_{t-i} \tag{2}$$

$$\text{where } r_i = \frac{\Gamma(i+d)}{\Gamma(i+1)\Gamma(d)} = \prod_{s=1}^{i} \frac{s+d-1}{s}$$

It is important to note that the above asymptotic expansion renders $X_t$ as a *linear combination of its innovations* $\in_{t-i}$, hence the above process is a linear process irrespective whether $\in_{t-i}$ is Gaussian or non-Gaussian. The parameter $d$ is related to the Hurst exponent (α) as $\alpha = d + 0.5$ and falls in the range $-0.5 < d < 0.5$ [12] for stationary FARIMA (0, d, 0). While short term correlations in the FARIMA (p, d, q) model are dictated by $f(B)$ and $y(B)$, long-range and anti-correlations are dictated by $d$. More importantly, one can realize long-range correlations for parameter $d$ in the range $(0 < d < 0.5)$. In order to minimize possible crossovers that are an outcome of short-range correlation we restrict the present study solely to FARIMA (0, d, 0).

In the subsequent sections we shall use the following abbreviations

**AWGN** (additive white Gaussian noise): FARIMA (0, d, 0), with innovations sampled from Gaussian distributed i.i.d process, normalized to zero-mean unit variance. By definition AWGN is a linear process.

**NAWGN** (nonlinear transform of AWGN): AWGN passed through a static, invertible , nonlinear filter $X_n \sqrt{|X_n|}$ . The choice of this specific transform is encouraged by a recent study [8]. By definition NAWGN is a nonlinear process where the nature of nonlinearity is a static.

**AWNGN** (additive white non-Gaussian noise): FARIMA (0, d, 0), with innovations sampled from square of Gaussian distributed i.i.d process, normalized to zero-mean unit variance. It should be noted that nonlinear transform of a Gaussian noise is not the only way to realize non-Gaussian innovations. By definition AWNGN is a linear process.

**NAWNGN** (nonlinear transform of AWNGN): AWNGN passed through a static, invertible, nonlinear filter $X_n \sqrt{|X_n|}$ [11]. By definition NAWNGN is a nonlinear process where the nature of nonlinearity is static.

In the present study, AWGN and NAWGN were generated with parameter $d = 0.3$, corresponding to long range correlated noise with Hurst exponent, $\alpha = 0.8$. The asymptotic expansion (2) was truncated to the first ($M = 500$) terms after initial experimentation. The number of data points was chosen sufficiently long ($t = 1…10^5$) after discarding the initial transients (N = 50000 samples). NAWGN and NAWNGN were generated as static, invertible nonlinear transforms ($X_n\sqrt{|X_n|}$) of AWGN and AWNGN respectively. Following [2-4], volatility series of AWGN, AWNGN, NAWGN and NAWNGN were generated as absolute value of their increments given by the expression $V_t = |X_t - X_{t-1}|$. Scaling of the volatility series was estimated using DFA with fourth order polynomial detrending [5].

*2.2 Surrogate Testing*

Surrogate testing is used widely to make statistical inference of the process generating the given single realization. Under implicit ergodic assumptions, this single realization or empirical sample is thought to be representative of the underlying dynamics. In the present study, we consider empirical samples generated from AWGN, AWNGN, NAWGN and NAWNGN as described in Sec. 2.1. Three essential ingredients of surrogate analysis include (a) null hypothesis (b) discriminant measure (c) algorithm to generate surrogates addressing that specific null hypothesis. Three popular surrogate algorithms used widely in literature include: phase-randomized surrogate (FT) [7], amplitude adjusted Fourier transform (AAFT) [8] and iterated amplitude adjusted Fourier transform (IAAFT) [8]. FT surrogates address the null that the given empirical sample is generated by a linearly correlated process with Gaussian innovations (e.g. AWGN). AAFT and IAAFT were designed to incorporate non-Gaussianity and address the null that the given empirical sample is generated by static, invertible nonlinear transform of a linear correlated noise. It might not be surprising to note that the null addressed by the FT surrogates is encapsulated by AAFT and IAAFT surrogates. Recent studies have pointed out the superiority of IAAFT compared to AAFT [8]. The details of the above surrogate algorithms can be found elsewhere [7, 8]. While we briefly discuss the impact of the three surrogate algorithms, the focus is on IAAFT surrogates. The scaling of the volatility series determined using DFA with fourth order polynomial detrending is used as a discriminant measure.

### 3. Results

As noted earlier, Sec. 2.1, AWGN and AWNGN were generated from FARIMA (0, d, 0) with d = 0.3 or $\alpha = 0.8$, M = 500, N = 100000 after discarding initial transients. NAWGN and NAWNGN were generated as static, invertible nonlinear transforms ($X_n \sqrt{|X_n|}$) of NAWGN and NAWNGN respectively. DFA with fourth order polynomial detrending was used to determine the scaling behavior of all the data sets in the present study.

The distribution of AWGN, NAWGN, AWNGN and NAWNGN is shown in Figs. 1a, 1b, 1c, and 1d respectively. The scaling exponent of AWGN and AWNGN with (d = 0.3) was $\alpha = 0.8$ and immune to the choice of the innovations (Gaussian or non-Gaussian), as expected, Fig. 1d. Estimating the scaling exponent of the empirical samples prior to analysis of their volatility series serves as a sanity check and justifies the finite sample size approximation ($M = 600$) of the asymptotic expansion (2) is not unreasonable.

By definition FT, IAAFT surrogate algorithms retain the power spectrum of the given empirical sample in the surrogate realization to a higher degree of accuracy than AAFT. As a preliminary check the power spectrum of the empirical samples and their surrogate counterparts were inspected. Retaining the power-spectrum implies retaining the two-point correlation, hence the scaling exponent. DFA exponent is related to the two-point correlation, thus we expect the scaling exponent estimates of AWGN, NAWGN, AWNGN, NAWNGN and their surrogate counterparts will not exhibit any significant difference. However, such a conclusion need not necessarily hold for their volatility series. Therefore, we restrict the following discussion solely to the scaling of the volatility series of AWGN, NAWGN, AWNGN and NAWNGN and their surrogate counterparts.

Power spectrum of the empirical samples generated by volatility transform of AWGN, NAWGN, AWNGN and NAWNGN and those of their corresponding FT, AAFT and IAAFT surrogate realizations is shown in Figs. 2a, 2b, 2c and 2d respectively.

**AWGN**: Monofractal noise generated from a linear process with Gaussian innovations, Sec. 2.1. Volatility series of AWGN and those of their FT, AAFT, IAAFT surrogate counterparts exhibited similar spectral decay characteristic of uncorrelated noise, $\alpha = 0.5$. Thus the null that the given data is a linearly correlated noise with Gaussian innovations cannot be rejected. These results conform to earlier studies on uncorrelated volatility correlations in monofractal noise.

**NAWGN**: Generated as nonlinear transform of linearly correlated monofractal noise (AWGN), Sec. 2.1, where the nonlinearity is a static invertible nonlinearity. Spectral decay of the volatility series of NAWGN exhibited considerable deviation from those of their FT surrogates. This is to be expected as the static, nonlinear transform is not retained by FT surrogates. Alternatively, the null that the given data is generated by a linearly correlated noise can be rejected. However, volatility series of NAWGN and those of their AAFT and IAAFT counterparts exhibited similar spectral decay. Thus the null that the given data is a static, invertible, nonlinear transform of a linearly correlated noise cannot be rejected. This can be attributed to the fact that AAFT and IAAFT surrogates by their very construction retain the static, invertible nonlinear transform, with IAAFT being superior to AAFT.

**AWNGN**: Monofractal noise generated from a linearly correlated process with non-Gaussian innovations, Sec. 2.1. Spectral decay of the volatility series of AWNGN exhibit considerable deviation from those of their AAFT and IAAFT counterparts. More importantly, volatility series of AAFT and IAAFT surrogates exhibit a dominant low-frequency signature unlike those of AWNGN. Therefore, the null that the given data is a static, nonlinear transform of a linearly correlated noise cannot be rejected and argues in favor of possible dynamical nonlinearity in the given data. Interestingly, spectral decay of the volatility series of FT surrogates is similar to those of AWNGN. Thus the null that the given data is a linearly correlated noise with Gaussian innovations cannot be rejected. These results are contrary to the usual norm, as the null hypotheses addressed by the AAFT and IAAFT surrogates encapsulate those addressed by FT surrogates. Alternatively, if the null is rejected by AAFT and IAAFT, it has to be rejected by FT. The failure to observe any discrepancies in the case of FT surrogates can be attributed to the fact that FT surrogates does not retain the phase information and implicitly renders the empirical sample (AWNGN) as a linearly correlated Gaussian noise. From previous reports and those of AWGN discussed earlier, volatility series of linearly correlated Gaussian noise resembles those of uncorrelated noise ($\alpha = 0.5$). The fact that FT surrogate is immune to the distribution of the empirical sample should discourage its use in non-Gaussian settings such as AWNGN.

**NAWNGN**: Generated as nonlinear transform of linearly correlated monofractal noise (AWNGN), by definition Sec. 2.1, where the nonlinearity is a static invertible nonlinearity. Spectral decay of volatility series of NAWGN exhibit considerable discrepancy from those of their FT, AAFT and IAAFT surrogate counterparts. Thus all three surrogate algorithms failed to reject the null arguing in favor of dynamical nonlinearity in the given empirical sample.

Similar investigation was carried out on the volatility series from AWGN, AWNGN, NAWGN, NAWNGN and their IAAFT surrogate counterparts using DFA with fourth order polynomial detrending, Figs. 3 and 4. Fifteen independent surrogate realizations were generated for each of the cases. Plot of the fluctuation function $\log_2 F(s)$ versus time scale $\log_2(s)$ is shown in fig. 3 whereas plot of the local slope $\alpha(s)$ versus timescale $\log_2(s)$ is shown in Fig. 4. Local slopes were estimated by linear regression of overlapping moving windows. This was accomplished by choosing a window containing ten points, estimate the exponent by local linear regression of the points in that window, shift the window by two points and repeat the procedure. Thus as a result, we obtain the scaling exponents $\alpha(s)$ as a function of the time scales $\log_2(s)$. A considerable discrepancy was observed in the volatility scaling of AWNGN, NAWNGN and their IAAFT surrogate counterparts. More importantly, volatility exponent of IAAFT surrogates is considerably larger than those of AWNGN and NAWNGN. This has to be contrasted with the volatility scaling of AWGN, NAWGN and their IAAFT counterparts which failed to exhibit any discrepancy.

Thus from the above case studies, Figs. 2, 3 and 4 it can be noted that volatility analysis in conjunction with FT, AAFT and IAAFT surrogate algorithms can give rise to false-positive identification of dynamical nonlinearity in the case of FARIMA (0, d, 0) with non-Gaussian innovations unlike those of Gaussian innovations. It is also clear that long-range volatility correlation due to static, invertible nonlinear transform can be accommodated by surrogate analysis in the case of FARIMA (0, d, 0) with Gaussian innovations, however, such an approach might not be adequate in the case of FARIMA (0, d, 0) with non-Gaussian innovations.

## 4. Discussion

Long-range volatility scaling has been found to be an indicator of nonlinear/multifractal dynamics in the given data. However, long-range volatility correlation can be an outcome of static as well as dynamical nonlinearity. Static nonlinearity such as those from measurement device has no relevance to the underlying dynamics, thus considered uninteresting. In order to make a finer distinction between dynamical and static nonlinearities, volatility analysis is used in conjunction with surrogate testing. In the present study, surrogate testing of volatility series generated from FARIMA (0, d, 0) processes with Gaussian and non-Gaussian innovations and their static, invertible nonlinear transforms were considered. We found that FT, AAFT and IAAFT surrogate algorithms were useful in statistical inference of FARIMA (0, d, 0) with Gaussian innovations (AWGN) and their static, invertible nonlinear transforms (NAWGN). However, volatility scaling of FARIMA (0, d, 0) with non-Gaussian innovations (AWNGN) and their static, invertible nonlinear transforms (NAWNGN) showed significant discrepancies from those of their AAFT and IAAFT counterparts. While volatility correlation of AWNGN exhibited uncorrelated behavior across all time scales, those of their AAFT and IAAFT surrogates of AWNGN exhibited considerable correlations across large time scales. Interestingly, volatility scaling of AWNGN failed to exhibit any significant change from their FT surrogate counterparts. This is anomalous since the null addressed by AAFT and IAAFT encapsulate those addressed by FT. However, in the present context the anomaly can be attributed to the fact that FT surrogates retain the two-point correlation and implicitly render the distribution of AWNGN normal. Volatility scaling of FT, AAFT and IAAFT surrogates of NAWNGN showed significant deviation from those of NAWNGN. While volatility analysis in conjunction with surrogate testing (FT, AAFT and IAAFT) can be useful in statistical inference of FARIMA (0, d, 0) with Gaussian innovations and their static, invertible nonlinear transforms, their choice in the presence of non-Gaussian innovations is limited. Alternatively, a significant difference in the volatility scaling between the given empirical sample and their surrogate counterpart can be solely due to non-Gaussianity with no regards to dynamical or even static nonlinearity.

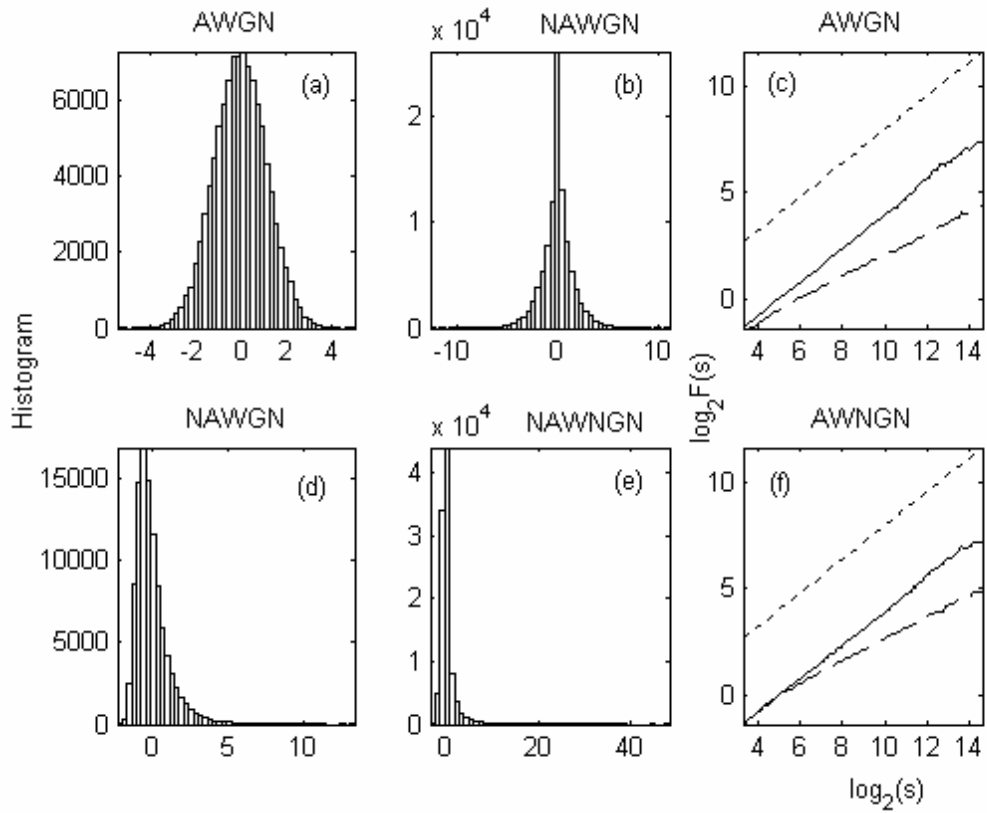

**Figure 1** Histogram of AWGN, NAWGN, AWNGN and NAWNGN ($N = 10^5$) is shown in (a, b, d and e) respectively. The fluctuation function $\log_2 F(s)$ versus time scale $\log_2(s)$ obtained using DFA with fourth order polynomial detrending for AWGN and AWNGN (solid line) and their corresponding volatility series (dashed line) is shown in (c and f) respectively. The dotted line in (c and f) is shown as a reference and corresponds to ($\alpha = 0.8$).

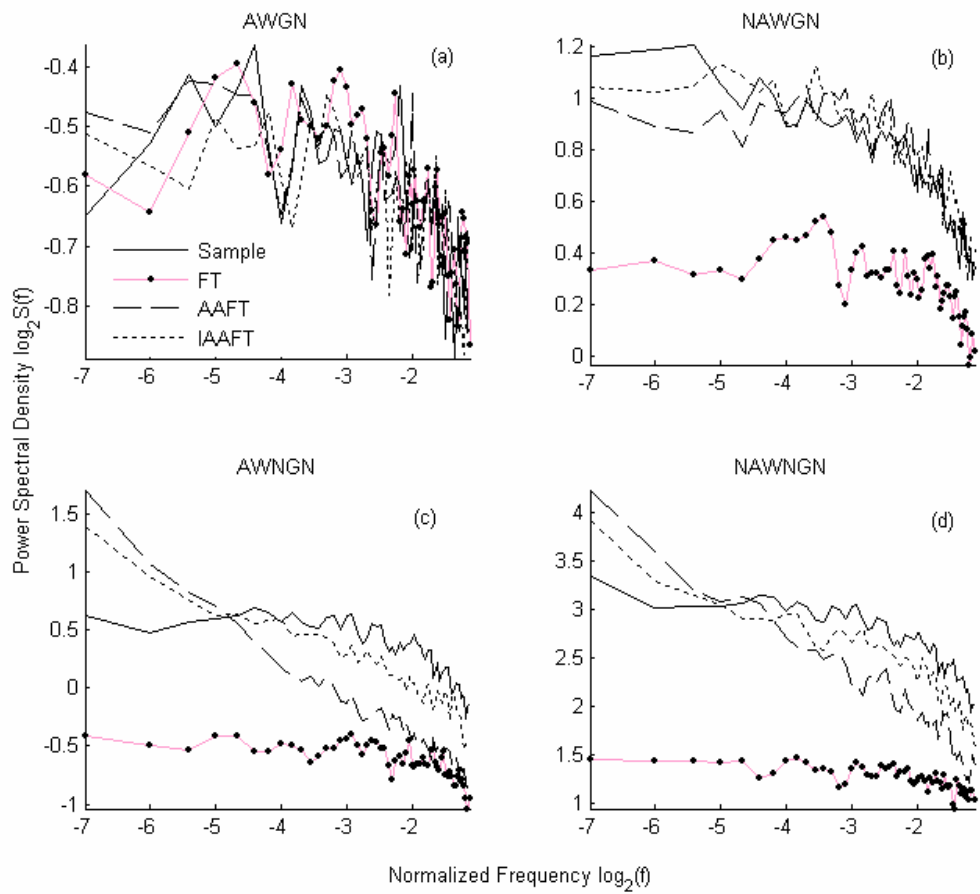

**Figure 2** Power spectral density $\log_2 S(f)$ versus normalized frequency $\log_2(f)$ of the volatility series ($N = 10^5$) of the empirical samples generated from AWGN, NAWGN, AWNGN, NAWNGN along with their corresponding FT, AAFT and IAAFT surrogate realizations is shown in (a, b, c and d) respectively. The legends for the subplots are identical and enclosed in (a), the term sample in the legend corresponds to empirical sample.

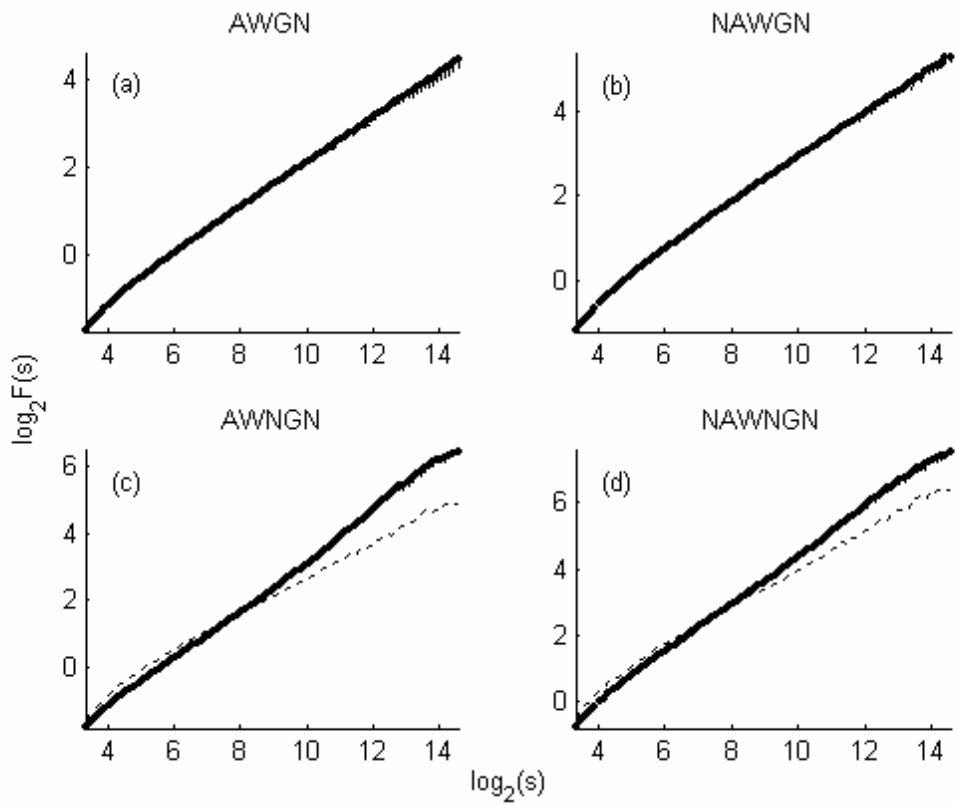

**Figure 3** Plot of the fluctuation function $\log_2 F(s)$ versus time scale $\log_2(s)$ for the volatility series ($N = 10^5$) of the empirical samples generated from AWGN, NAWGN, AWNGN, NAWNGN (thin dotted lines) along with their corresponding IAAFT surrogate counterparts (solid lines) is shown in (a, b, c and d) respectively. The vertical lines correspond to standard deviation about the mean value for 15 independent IAAFT surrogate realizations.

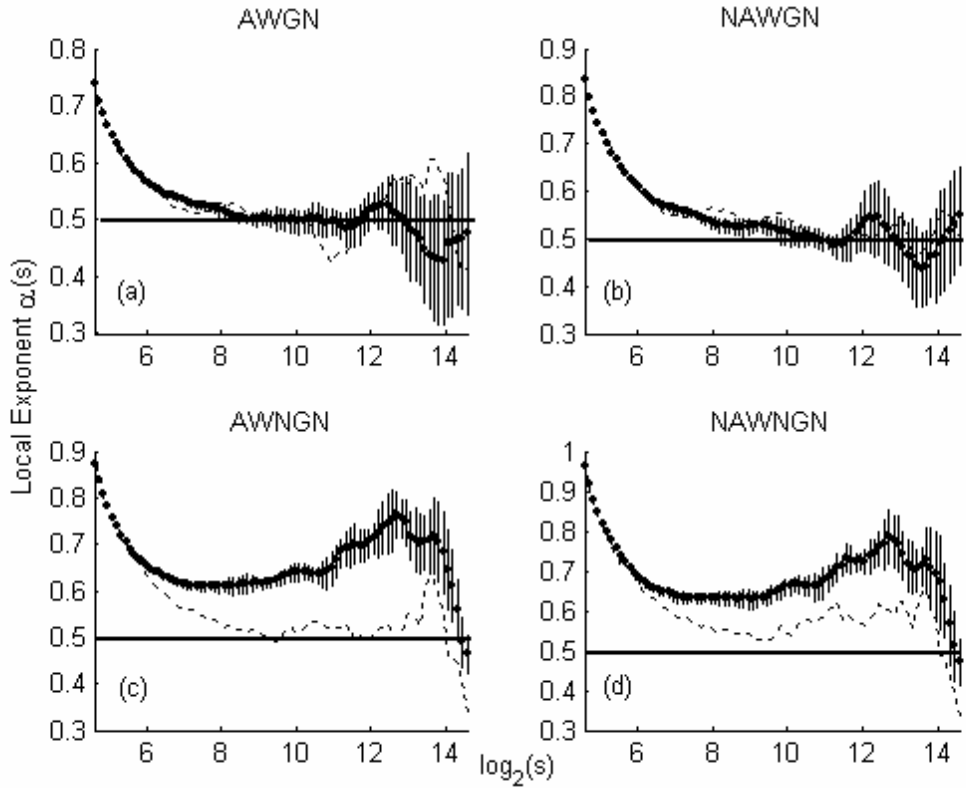

**Figure 4** Plot of the local exponent α(s) versus time scale log$_2$(s) for the volatility series (N = 10$^5$) of the empirical samples generated from AWGN, NAWGN, AWNGN, NAWNGN (thin dotted lines) along with their corresponding IAAFT surrogate counterparts (solid lines) is shown in (a, b, c and d) respectively. The vertical lines correspond to standard deviation about the mean value for 15 independent IAAFT surrogate realizations.